\documentclass[aps,prc,twocolumn,preprintnumbers,amsmath,amssymb,superscriptaddress,floatfix,longbibliography]{revtex4-2}
\usepackage{graphicx}
\usepackage{dcolumn}
\usepackage{bm}
\usepackage{color} 

\usepackage[colorlinks=true, linkcolor=blue, citecolor=blue, urlcolor=blue]{hyperref}

\def\lesssim{\ \raise.3ex\hbox{$<$}\kern-0.8em\lower.7ex\hbox{$\sim$}\ }
\def\gesim{\ \raise.3ex\hbox{$>$}\kern-0.8em\lower.7ex\hbox{$\sim$}\ }

\newcommand \beq{\begin{eqnarray}}
\newcommand \eeq{\end{eqnarray}}

\newcommand{\red}[1]{{#1}}

\begin{document}
\preprint{RIKEN-iTHEMS-Report-24}
\preprint{NITEP 221}

\title{Polaronic neutron in dilute alpha matter: A $p$-wave Bose polaron}
\author{Hiroyuki Tajima}
\affiliation{Department of Physics, Graduate School of Science, The University of Tokyo, Tokyo 113-0033, Japan}
\affiliation{RIKEN Nishina Center, Wako 351-0198, Japan}
\author{Hajime Moriya}
\affiliation{Research Center for Nuclear Physics, Osaka University, Ibaraki, Osaka 567-0047, Japan}
\author{Tomoya Naito}
\affiliation{Interdisciplinary Theoretical and Mathematical Science Program (iTHEMS), RIKEN, Saitama 351-0198, Japan} 
\affiliation{Department of Physics, Graduate School of Science, The University of Tokyo, Tokyo 113-0033, Japan}
\author{Wataru Horiuchi}
\affiliation{Department of Physics, Osaka Metropolitan University, Osaka 558-8585, Japan}
\affiliation{Nambu Yoichiro Institute of Theoretical and Experimental Physics (NITEP), Osaka Metropolitan University, Osaka 558-8585, Japan}
\affiliation{RIKEN Nishina Center, Wako 351-0198, Japan}
\affiliation{Department of Physics, Hokkaido University, Sapporo 060-0810,
  Japan}
\author{Eiji Nakano}
\affiliation{Department of Mathematics and Physics, Kochi University, Kochi 780-8520, Japan}
\author{Kei Iida}
\affiliation{Department of Mathematics and Physics, Kochi University, Kochi 780-8520, Japan}
\affiliation{RIKEN Nishina Center, Wako 351-0198, Japan}

\begin{abstract}
We theoretically investigate quasiparticle properties of a neutron immersed in an alpha condensate, which is
one of the possible states of dilute symmetric nuclear matter.
The resonant $p$-wave neutron-alpha scattering, which plays a crucial role in forming halo nuclei, is considered.
This system is similar to a Bose polaron near the $p$-wave Feshbach resonance that can be realized in cold-atomic experiments.
Calculating the self-energy within the field-theoretical approach, we give an analytical formula for the effective mass of a polaronic neutron as a function of alpha condensation density.
Moreover, two adjacent neutrons in a medium, each of which behaves like a stable polaron having an enhanced effective mass,
can form a bound dineutron,
with the help of $^1S_0$ neutron-neutron attraction.
This is in contrast to the case of the vacuum,  where a dineutron is known to be unbound.
Our result would be useful for understanding many-body physics in astrophysical environments as well as the formation of multi-nucleon clusters in neutron-halo nuclei.
\end{abstract}
\maketitle
\section{Introduction}
Infinite symmetric nuclear matter, where the proton charge is switched off, is fictitious, but is one of the most fundamental nuclear systems because it is related to saturation of the binding energy and density of stable nuclei~\cite{BLAIZOT1980171}.
It is still interesting to consider 
infinite symmetric nuclear matter at relatively low densities, although naively the system 
undergoes an instability towards inhomogeneous phases
at sufficiently low temperature~\cite{PhysRevLett.61.818}.
Once the proton charge is switched on, the nuclear liquid has to be a collection of nuclei.  Furthermore, such symmetric matter can contain alpha particles via alpha clustering; indeed, some medium-heavy symmetric nuclei 
are predicted to be unstable with respect to alpha decay~\cite{RevModPhys.84.567}. 

The nuclear equation of state and the composition in nuclear statistical equilibrium are important keys to understanding 
the mechanism of supernova explosions~\cite{RevModPhys.89.015007} and the dynamics of intermediate-energy heavy-ion collisions.
To this end, dilute alpha matter and its Bose-Einstein condensation (BEC) phase have been studied as a good reference system for supernova matter at finite temperatures~\cite{PhysRevC.99.024909,PhysRevC.101.024913,FURUSAWA2020121991,PhysRevC.103.024301}.
Alpha particles in such an astrophysical environment play a pivotal role in nucleosynthesis, e.g., triple-alpha reaction via the Hoyle state~\cite{hoyle1954resonances}. Moreover, the alpha condensates have also been discussed in the context of excited states of atomic nuclei~\cite{PhysRevLett.87.192501,PhysRevLett.101.082502}, which are expected to be present in intermediate energy heavy-ion collisions.
However, 
the many-nucleon and dynamical nature of the systems of interest here leads to a lot of uncertainties and model dependencies.

To explore the aforementioned alpha clustering properties more clearly,
various low-energy nuclear experiments have been performed.
The existence of the alpha condensates has been examined experimentally such as $^{16}$O~\cite{WAKASA2007173} and $^{20}$Ne~\cite{ADACHI2021136411}.
In addition, experimental explorations of multi-neutron clusters in nuclei of large neutron excess are ongoing~\cite{PhysRevLett.116.052501,FAESTERMANN2022136799,duer2022observation,PhysRevLett.133.012501}.
While the existence of the tetraneutron resonance has been discussed based on few-body calculations with multi-nucleon interactions~\cite{PhysRevLett.117.182502,PhysRevLett.118.232501,PhysRevLett.130.102501} (see also a review~\cite{marques2021quest}),
it is known that the neutron-nucleus interaction plays a crucial role in the formation of halo nuclei~\cite{Hammer_2017}.
There are several examples of two-neutron halo nuclei, $^6$He~\cite{PhysRevLett.55.2676}, $^{11}$Li~\cite{PhysRevLett.100.192502},
$^{14}$Be~\cite{PhysRevLett.86.600},
$^{19}$B~\cite{PhysRevLett.124.212503},
$^{22}$C~\cite{PhysRevC.74.034311,PhysRevLett.104.062701,TOGANO2016412,PhysRevC.97.054614},
and $^{29}$F~\cite{PhysRevC.101.024310,PhysRevLett.124.222504}.
Moreover, a recent experiment indicates that
dineutron correlations play a crucial role in understanding the cluster structure of $^8$He~\cite{PhysRevLett.131.242501,PhysRevC.108.L011304}.
In between, neutrons and alpha clusters are expected to coexist; alpha clusters in a neutron-skin region of stable tin isotopes with small neutron excess have just recently been identified by a knockout experiment~\cite{tanaka2021formation}.

\begin{figure}[t]
    \centering
    \includegraphics[width=6cm]{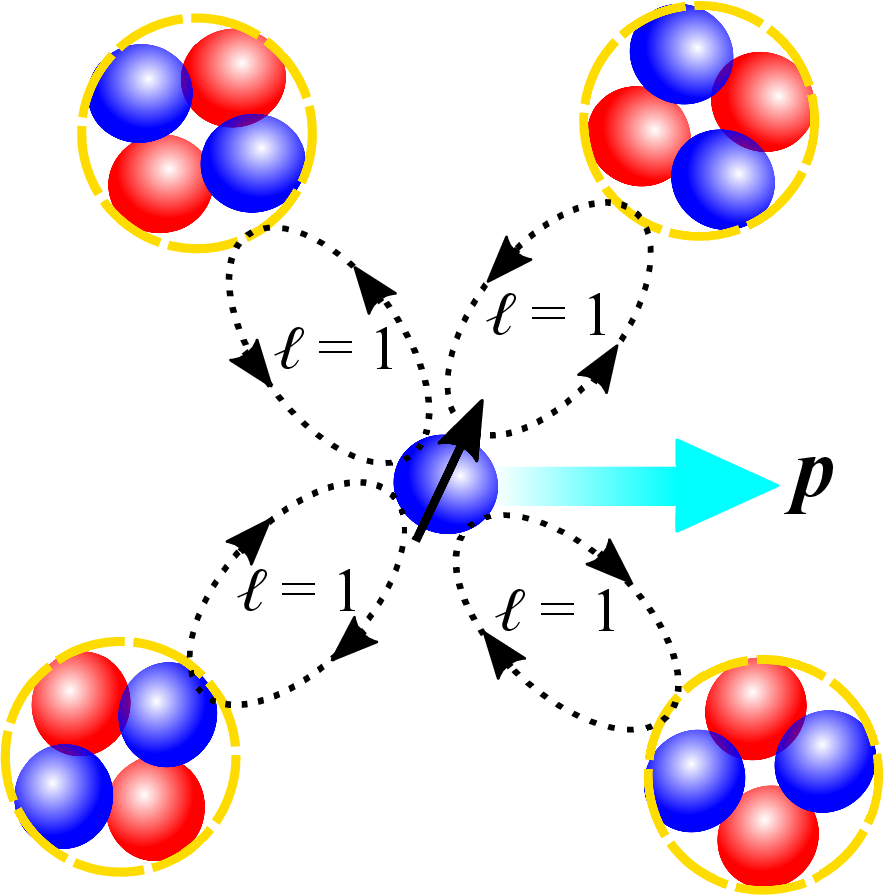}
    \caption{Schematic picture of a polaronic neutron moving with momentum $\red{\bm{p}}$ in an alpha condensate, which is characterized by the neutron-alpha $J^\pi=3/2^{-}$ $p$-wave resonance (i.e., relative angular momentum $\ell=1$).} 
    \label{fig:1}
\end{figure}

Generally, to understand how the properties of an impurity particle change in medium from those in vacuum,   
the notion of a polaron, which was originally proposed to see the properties of mobile electrons in ionic lattices~\cite{landau1933electron,landau1948effective}, is useful. This is the case with nuclear systems in which the medium and impurity particles can be single nucleons or nucleon composites (nuclear clusters).
Polarons are now one of the hot topics in cold atomic physics because of its tunable setup in recent experiments~\cite{baroni2024quantum}.
Some of the present authors clarified the quasiparticle properties of impurity protons and alpha particles
in dilute neutron matter as encountered in the crust of a neutron star, in terms of
Fermi polarons (i.e., impurities immersed in a Fermi sea)~\cite{PhysRevC.102.055802,PhysRevC.104.065801,tajima2024intersections,TAJIMA2024138567}.
In this regard, it is interesting to consider how neutrons are affected by alpha matter as shown in Fig.~\ref{fig:1}, which can be relevant to intermediate energy heavy-ion collisions, stellar collapse, and the neutron-skin region of atomic nuclei.
Indeed, a similar situation, that is, an impurity surrounded by BEC, is realized in cold-atomic systems as a so-called Bose polaron.
Accordingly, the microscopic properties of Bose polarons have been extensively studied experimentally~\cite{PhysRevLett.117.055301,PhysRevLett.117.055302,yan2020bose,duda2023transition} and theoretically~\cite{PhysRevA.88.053632,PhysRevLett.115.125302,PhysRevA.96.013607,PhysRevA.99.063607,PhysRevA.100.023624,PhysRevLett.122.183001,PhysRevLett.120.050405,PhysRevA.104.023317,atoms9010018} (see also a review~\cite{atoms10020055}).

In this paper, we theoretically investigate the quasiparticle properties of a neutron immersed in the alpha condensation state of dilute symmetric nuclear matter,  
as schematically depicted in Fig.~\ref{fig:1}.
We focus on the resonant $J^\pi=3/2^{-}$ $p$-wave neutron-alpha coupling, which 
is responsible for the halo structure of neutron-rich nuclei such as $^6$He~\cite{Hammer_2017}.
In this regard, the system we are interested in is similar to Bose polarons near the $p$-wave Feshbach resonance,
which would be experimentally accessible in several Bose-Fermi mixtures (e.g., $^6$Li-$^{87}$Rb~\cite{PhysRevA.78.022710},
$^{40}$K-$^{41}$K~\cite{PhysRevA.84.011601},
$^{23}$Na-$^{40}$K~\cite{PhysRevA.85.051602}, and $^6$Li-$^{133}$Cs~\cite{PhysRevA.87.010701}).
This system is different from the one treated in the previous work on polaronic neutrons in spin-polarized neutron matter~\cite{PhysRevC.89.041301,PhysRevC.103.L052801,PhysRevC.108.L052802}.
While $p$-wave Fermi polarons have been investigated theoretically~\cite{PhysRevLett.109.075302,PhysRevA.100.062712},
its bosonic counterpart has yet to be addressed.
It is still noteworthy that
recently, the role of interspecies $p$-wave interaction has been discussed in the context of a magnon in a two-dimensional Bose-Bose mixture~\cite{PhysRevB.109.235135}.
Here, we analyze the polaronic neutron excitation
by calculating the self-energy for the resonant $p$-wave neutron-alpha scattering.
Furthermore, by considering the residual $^1S_0$ neutron-neutron attractive interaction, we study the fate of two adjacent polaronic neutrons in dilute alpha matter.  
The possible occurrence of bound dineutrons due to the polaron effect, which is explored by solving the Bethe-Salpeter equation,
may be relevant to the formation of multi-nucleon clusters in neutron-rich nuclei, intermediate energy heavy-ion collisions, and core-collapse supernova cores.
Moreover, in the presence of such a bound state, the crossover from dineutron BEC to Bardeen-Cooper-Schrieffer (BCS) neutron superfluid can be anticipated~\cite{PhysRevC.73.044309}.

This paper is organized as follows.
In Sec.~\ref{sec:2}, the theoretical model and the Bose-polaron formalism will be presented.
In Sec.~\ref{sec:3},
we will show the result for the polaronic neutron excitations and therefrom discuss the dineutron formation in the alpha condensate.
Summary of this paper will be shown in Sec.~\ref{sec:4}.

\section{Formalism}
\label{sec:2}
Hereinafter, we take $\hbar=k_{\rm B}=1$, and the system volume is set to unity.
\subsection{Model}
Here, we employ the two-channel Hamiltonian~\cite{PhysRevC.106.045807} for the $J^{\pi}=3/2^{-}$ $^5$He resonance, which is relevant at low relative energies, as given by
\begin{equation}
    H=H_{\nu}+H_{\alpha}+H_{\Phi}+V_{3/2^-},
\end{equation}
where the contribution of spin-$1/2$ neutrons ($s_z=\pm 1/2$) with a mass $M_\nu$ reads
\begin{align}
  H_{\nu}
  & =
    \sum_{s_z} \sum_{\bm{k}}
    \xi_{\bm{k},\nu}
    \nu_{\bm{k},s_z}^\dag \nu_{\bm{k},s_z} \cr
  & \quad
    +
    \sum_{\bm{k},\bm{k}',\bm{Q}}U_{2\nu}(\bm{k},\bm{k}')
    \nu_{\bm{k}+\bm{Q}/2,+1/2}^\dag
    \nu_{-\bm{k}+\bm{Q}/2,-1/2}^\dag \cr
  &\qquad
    \times \nu_{-\bm{k}'+\bm{Q}/2,-1/2}
    \nu_{\bm{k}'+\bm{Q}/2,+1/2}
\end{align}
with the neutron kinetic energy $\xi_{\bm{k},\nu}=k^2/2M_\nu$,
the neutron creation (annihilation) operator $\nu_{\bm{k},s_z}^{(\dag)}$,
and $^1S_0$ neutron-neutron coupling $U_{2\nu}(\bm{k},\bm{k}')$.
The contribution of alpha particles reads
\begin{align}
  H_{\alpha}
  & =
    \sum_{\bm{q}}\xi_{\bm{q},\alpha}\alpha_{\bm{q}}^\dag \alpha_{\bm{q}}\cr
  & \quad
    +\frac{1}{2}\sum_{\bm{q},\bm{q}',\bm{K}}
    U_{2\alpha}(\bm{q},\bm{q}')
    \alpha_{\bm{q}+\bm{K}/2}^\dag
    \alpha_{-\bm{q}+\bm{K}/2}^\dag\cr
  &\qquad
    \times
    \alpha_{-\bm{q}'+\bm{K}/2}
    \alpha_{\bm{q}'+\bm{K}/2}
\end{align}
with the alpha-particle kinetic energy $\xi_{\bm{q},\alpha}=q^2/2M_\alpha-\mu_\alpha$ ($M_\alpha=4M_\nu$ and $\mu_\alpha$ are the alpha particle mass and chemical potential, respectively),
the alpha-particle annihilation (creation) operator $\alpha_{\bm{q}}^{(\dag)}$,
and the alpha-alpha interaction $U_{2\alpha}(\bm{q},\bm{q}')$,
and the kinetic term of the closed-channel $^5$He state with $J^\pi=3/2^-$ ($J_z=\pm 1/2 $, $ \pm 3/2$) reads
\begin{align}
    H_{\Phi}
    =\sum_{\bm{P},J_z}\left(\xi_{\bm{P},\Phi}+E_{\Phi}\right)
    \Phi_{\bm{P},J_z}^\dag \Phi_{\bm{P},J_z}
\end{align}
with the bare $^5$He kinetic energy $\xi_{\bm{P},\Phi}=P^2/\red{(2M_\nu+2M_\alpha)}-\mu_\alpha$,
and the bare $^5$He energy level $E_{\Phi}$,
the 
bare $^5$He annihilation (creation) operator $\Phi_{\bm{P},J_z}^{(\dag)}$.
Note that we set the neutron chemical potential to zero since neutrons are regarded as impurities.

\red{For a two-body system of a neutron with momentum $\bm{p}_\nu$ and an alpha particle with momentum $\bm{p}_\alpha$,
using the center-of-mass momentum $\bm{P}=\bm{p}_\nu+\bm{p}_\alpha$ and
the relative momentum $\bm{k}=M_{\rm r}\left(\frac{\bm{p}_\nu}{M_\nu}-\frac{\bm{p}_\alpha}{M_\alpha}\right)$ with the reduced mass $M_{\rm r}^{-1}=M_\nu^{-1}+M_{\alpha}^{-1}$,
one can rewrite $\bm{p}_\nu$
and $\bm{p}_\alpha$ as
\begin{align}
    \bm{p}_\nu
    &=\bm{k}+\frac{M_\nu}{M_\nu+M_\alpha}\bm{P}
    \equiv \bm{k}+\frac{\bm{P}}{5}, \\
    \bm{p}_\alpha
    &=-\bm{k}+\frac{M_\alpha}{M_\nu+M_\alpha}\bm{P}
    \equiv -\bm{k}+\frac{4\bm{P}}{5},
\label{eq:p_alpha}
\end{align}
respectively.
}
Then, the interaction term $V_{3/2^-}$ responsible for the $J^\pi=3/2^{-}$ $p$-wave resonance is given by
\begin{widetext}    
\begin{equation}
    V_{3/2^-}=
    \sum_{J_z,s_z,m}
    \sum_{\bm{P},\bm{k}}
    \left(k{g}_{k}\sqrt{\frac{4\pi}{3}}Y_{m}^{\ell=1}(\hat{\bm{k}})
    \langle 1,m;1/2,s_z |3/2,J_z\rangle
    \Phi_{\bm{P},J_z}^\dag
    \red{\nu_{\bm{k}+\bm{P}/5,s_z}\alpha_{-\bm{k}+4\bm{P}/5}}+{\rm h.c.}\right),
\end{equation}
\end{widetext}
where $\langle 1,m;1/2,s_z |3/2,J_z\rangle$ is the Clebsch-Gordan coefficient
~\footnote{
\red{In Ref.~\cite{PhysRevC.106.045807}, the neutron and alpha-particle momenta in the center-of-mass frame were wrongly taken as $\bm{k}+\bm{P}/2$ and $-\bm{k}+\bm{P}/2$, respectively. Since only the case with $\bm{P}=\bm{0}$ was considered in Ref.~\cite{PhysRevC.106.045807}, however, the results were not affected by this choice at all.}}. 
We employ the Yamaguchi-type form factor
\begin{equation}
\label{eq:gamma}
    g_{k}=\frac{g}{1+(k/\Lambda)^2},
\end{equation}
where $\Lambda$ is the cutoff scale and $g$ is the coupling strength at $k=0$.
These parameters are related to the low-energy constants as~\cite{PhysRevC.106.045807}
\begin{align}
  a_p
  & =
    -\frac{M_{\rm r}g^2}{6\pi}\left(E_{\Phi}-\frac{M_{\rm r}g^2\Lambda^3}{12\pi}\right)^{-1},
    \notag \\
  r_p
  & =
    -\frac{6\pi}{M_{\rm r}^2g^2}+\frac{24\pi E_{\Phi}}{M_{\rm r}g^2\Lambda^2}-3\Lambda.
\end{align}
Here, $a_p=-67.1 \, \mathrm{fm}^3 $ and $r_p=-0.87 \, \mathrm{fm}^{-1}$ are the $J^{\pi}=3/2^{-}$ $p$-wave scattering volume and effective range, respectively~\cite{PhysRevC.102.055802}
.
Also, we set $\Lambda=0.90 \, \mathrm{fm}^{-1}$ in such a way as to reproduce the $^5$He resonance energy $E_{\rm res}\sim 0.93 \, \mathrm{MeV} $. 
We can then determine $E_{\Phi}=449.607 \, \mathrm{MeV} $ and $M_{\rm r}g^2=113.388 \, \mathrm{fm}^2$ from the empirical values of $a_p$ and $r_p$.

\subsection{Bose-polaron self-energy}
As for a medium of dilute symmetric nuclear matter, we consider a zero-temperature alpha condensate.
The alpha condensate density denoted by $\rho_\alpha$ controls the polaronic properties of an impurity neutron. 
Within the mean-field approximation, the alpha condensation energy density reads $\mathcal{E}_{\alpha}=-\mu_\alpha\rho_\alpha+\frac{1}{2}U_{2\alpha}(\bm{0},\bm{0})\rho_\alpha^2$,
from which we obtain $\mu_\alpha=U_{2\alpha}(\bm{0},\bm{0})\rho_\alpha$.
However, since the value of $U_{2\alpha}(\bm{0},\bm{0})$ has a large uncertainty in the matter,
for simplicity, we take $\mu_\alpha\simeq 0$,
which is valid in the dilute regime.
In this regard,
the information on the alpha condensate is incorporated only via $\rho_\alpha$.
\begin{figure}[t]
    \centering
    \includegraphics[width=5cm]{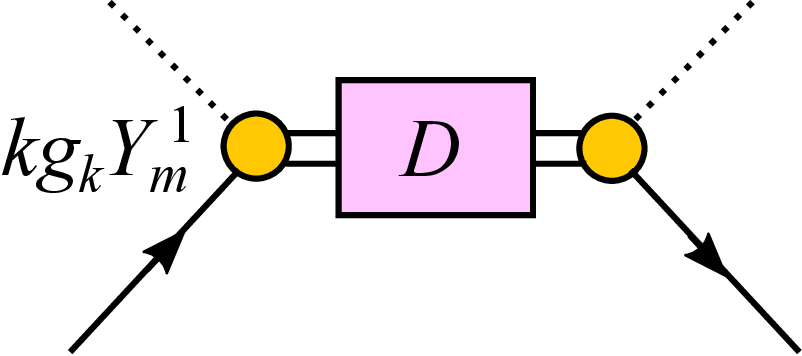}
    \caption{Beliaev-type self-energy of a polaronic neutron in an alpha condensate,  which is accompanied by the dressed $^5$He propagator $D$. The dotted line describes the 
    alpha condensate wave function or, equivalently, the square root of the condensate density $\sqrt{\rho_\alpha}$. The circles represent the neutron-alpha-$^5$He coupling $kg_kY_{m}^{1}$ within the $J^{\pi}=3/2^{-}$ channel.}
    \label{fig:2}
\end{figure}

We are interested in the neutron retarded Green's function
\begin{equation}
    G_{s_z}(\red{\bm{p}},\omega)=\frac{1}{\omega_+-\xi_{\red{\bm{p}},\nu}-\Sigma_{s_z}(\red{\bm{p}},\omega)},
\end{equation}
where 
$\omega_{+}=\omega+i\delta$ involves an infinitesimally small imaginary number $i\delta$.
Then, the self-energy of a neutron embedded in the alpha condensate is given by the Beliaev-type diagram~\cite{PhysRevA.88.053632}:
\begin{align}
    \Sigma_{s_z}(\red{\bm{p}},\omega)&=\frac{4\pi}{3}\rho_\alpha
    \sum_{J_z}\sum_{m}
    \langle 1,m;1/2,s_z|3/2,J_z\rangle^2\cr
    &
    \quad
    \times
    k^2g_{k}^2
    Y_{m}^{1}(\hat{\bm{k}})
    [Y_{m}^{1}(\hat{\bm{k}})]^{*}
    D_{J_z}(\red{\bm{p}},\omega),
\end{align}
which incorporates the resonant scattering between a neutron with \red{momentum $\bm{p}= \bm{p}_\nu=\bm{k}+\bm{P}/5$ and an alpha particle with zero momentum $\bm{p}_\alpha=-\bm{k}+4\bm{P}/5=\bm{0}$ 
(i.e., in the condensate).
In such a case, the relative momentum is given by
\begin{align}
    \bm{k}=
    \frac{1}{1+\frac{M_\nu}{M_\alpha}}\bm{p}
    =\frac{4}{5}\bm{p}.
\end{align}}
More explicitly, in the case of $s_z=+1/2$, we obtain
\begin{align}
    \Sigma_{1/2}(\red{\bm{p}},\omega)
    &=
    g_{k}^2\rho_\alpha
    \left[\frac{k_x^2+k_y^2}{2}
    D_{3/2}(\red{\bm{p}},\omega)\right. \cr
    &\quad\quad \left.   +\frac{k_x^2+k_y^2+4k_z^2}{6}
    D_{1/2}(\red{\bm{p}},\omega)
    \right], 
\end{align}
where the dressed $^5$He propagators $D_{J_z}(\red{\bm{p}},\omega)$ are given by
\begin{equation}
    D_{J_z}(\red{\bm{p}},\omega)
    =\frac{1}{\omega_+-\xi_{\red{\bm{p}},\Phi}-E_{\Phi}-\Pi_{J_z}(\red{\bm{p}},\omega)}.
\end{equation}
Note that $\Sigma_{s_z=-1/2}(\red{\bm{p}},\omega)=\Sigma_{s_z=1/2}(\red{\bm{p}},\omega)$.
The one-loop self-energy $\Pi_{J_z}(\bm{P},\omega)$ for the neutron-alpha scattering
reads
\begin{align}
\label{eq:pi3o2}
  \Pi_{3/2}(\bm{P},\Omega)
  & =
    \sum_{\bm{q}}g_{q}^2\left(\frac{q_x^2+q_y^2}{2}\right)\cr
  & \qquad
    \times\frac{1}{\Omega_+-\xi_{\red{\bm{q}+\bm{P}/5},\nu}-\xi_{\red{-\bm{q}+4\bm{P}/5},\alpha}}, \\
  \Pi_{1/2}(\bm{P},\Omega)
  & =
    \sum_{\bm{q}}g_{q}^2
    \left(\frac{2}{3}q_z^2+\frac{q_x^2+q_y^2}{6}\right)\cr
  &\qquad
    \times \frac{1}{\Omega_+-\xi_{\red{\bm{q}+\bm{P}/5},\nu}-\xi_{\red{-\bm{q}+4\bm{P}/5},\alpha}}.
\end{align}
\red{Using $\xi_{\red{\bm{q}+\bm{P}/5},\nu}+\xi_{\red{-\bm{q}+4\bm{P}/5},\alpha}=q^2/2M_{\rm r}+P^2/(2M_\nu+2M_\alpha)$,
we obtain 
$\Pi_{3/2}(\bm{P},\Omega)=\Pi_{1/2}(\bm{P},\Omega)=\Pi_0(\Omega_P)$ as
\begin{equation}
\label{eq:pi}
    \Pi_{0}(\Omega_P)
    =-
    \frac{M_{\rm r}g^2\Lambda^4[\Lambda^3+6\Lambda M_{\rm r}\Omega_P+2i(2M_{\rm r}\Omega_P)^{3/2}]}{12\pi(2M_{\rm r}\Omega_P+\Lambda^2)^2},
\end{equation}
with $\Omega_P=\Omega-P^2/(2M_\nu+2M_\alpha)$.
To obtain Eq.~\eqref{eq:pi},
we used $\sum_{\bm{q}}q_j^2F(q)=\frac{1}{3}\sum_{\bm{q}}q^2F(q)$ ($j=x,y,z$) for an arbitrary function $F(q=|\bm{q}|)$.
Eventually, 
with $\omega_p=\omega-p^2/(2M_\nu+2M_\alpha)$,
we find
\begin{align}
    \Sigma_{s_z}(\bm{p},\omega)
    =\frac{2}{3}
    \frac{k^2g_k^2\rho_\alpha}{\omega_+-\xi_{\bm{p},\Phi}-E_\Phi-\Pi_0(\omega_p)},
\end{align}
which is found to be isotropic in the momentum space,
in contrast to Fermi polarons with anisotropic interactions~\cite{PhysRevLett.109.075302,PhysRevA.103.033324}.
}

\section{Results}

Let us now exhibit numerical results for the properties of a Bose-polaronic neutron and the
resultant implications for the bound dineutron.

\label{sec:3}
\subsection{Polaronic neutron in the alpha condensate}
\begin{figure*}[t]
    \centering
    \includegraphics[width=14cm]{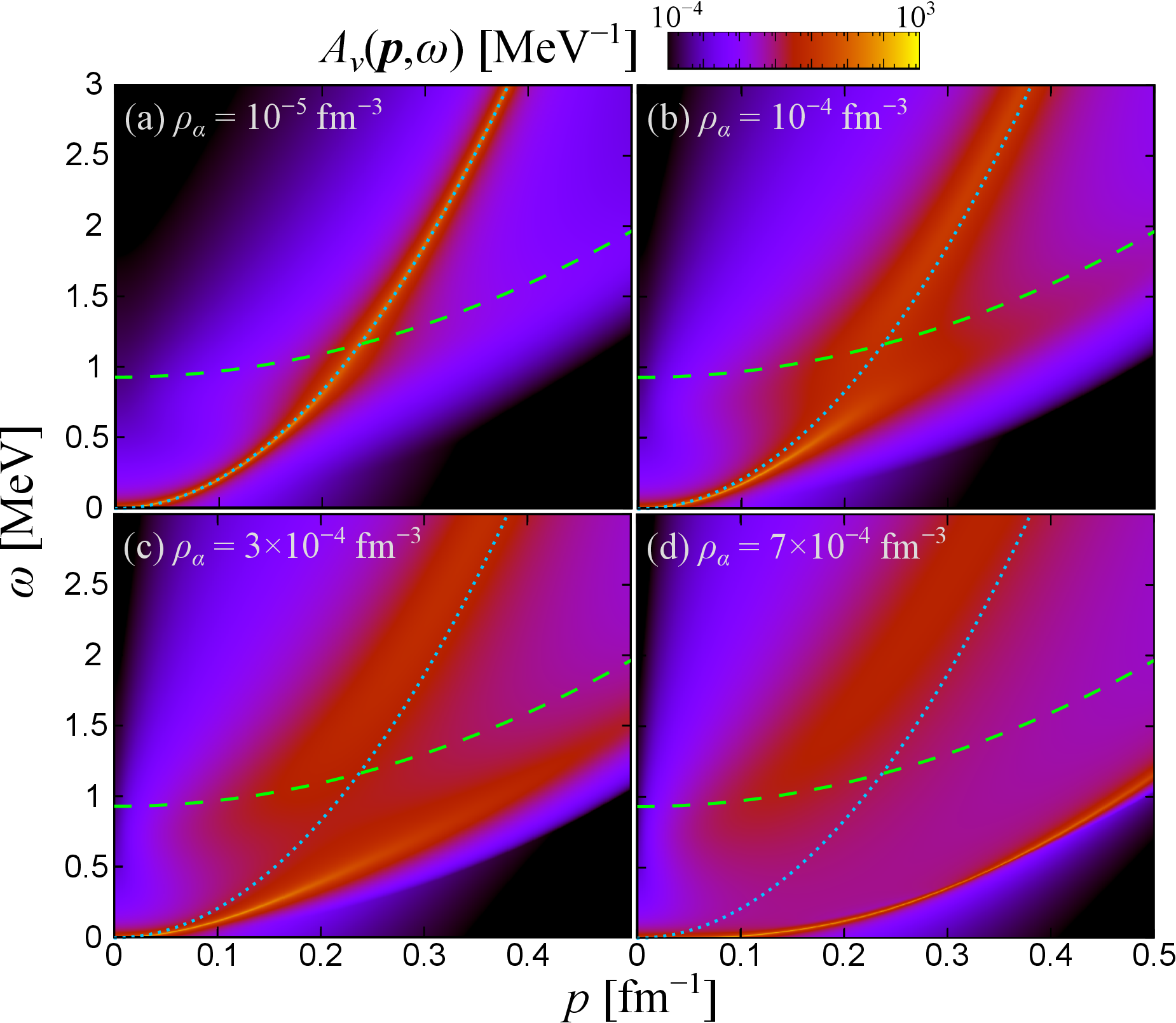}
    \caption{Polaronic neutron spectral weight $A_{\red{\nu}}(\red{\bm{p}},\omega)$ in the alpha condensate.
    For comparison, the bare neutron dispersion $\xi_{\red{\bm{p}},\nu}=\red{p}^2/2M_\nu$ (dotted curve)
    and $^5$He resonance branch $p^2/(2M_\nu+2M_\alpha)+E_{\rm res}$ (dashed curve)
    are plotted in each panel.  
      The alpha condensation densities are taken as
      (a) $\rho_\alpha=10^{-5} \, \mathrm{fm}^{-3}$,
      (b) $\rho_\alpha=10^{-4} \, \mathrm{fm}^{-3}$,
      (c) $\rho_\alpha=3\times 10^{-4} \, \mathrm{fm}^{-3}$,
      and
      (d) $\rho_\alpha=7\times 10^{-4} \, \mathrm{fm}^{-3}$.
    }
    \label{fig:3}
\end{figure*}
Figure~\ref{fig:3} shows the polaronic neutron spectral weight given by
\begin{align}
    A_{\red{\nu}}(\red{\bm{p}},\omega)=-\frac{1}{\pi}{\rm Im}G_{s_z}(\red{\bm{p}},\omega),
\end{align}
where we omitted the spin index \red{of $A_{\red{\nu}}(\red{\bm{p}},\omega)$}.
For the overall structure, one can see that a sharp peak remains typically up to the threshold momentum of $^5$He given by $\sqrt{2M_\nu E_{\rm res}}\simeq 0.21 \ {\rm fm}^{-1}$ in the low-density regime.
In the low-density regime, e.g., at $\rho_\alpha=10^{-5} \, \mathrm{fm}^{-3}$ in Fig.~\ref{fig:3}~(a),
one can see that the quasiparticle peak is located close to the bare dispersion $\xi_{\red{\bm{p}},\nu}=\red{p}^2/2M_\nu$ and that
the spectral broadening occurs near the $^5$He resonance. 
At $\rho_\alpha=10^{-4} \, \mathrm{fm}^{-3}$, as shown in Fig.~\ref{fig:3}~(b),
there is a deviation between the spectral peak and the bare dispersion in such a way that the effective mass of the polaronic neutron branch becomes larger than the bare one.
\red{This can be understood as the avoided-crossing behavior of neutron quasiparticle and $^5$He resonance branches.}
Moreover, the spectral broadening due to the coupling with the $^5$He resonance branch is more prominent at larger alpha condensate densities, i.e., $\rho_\alpha=3\times 10^{-4}$~fm$^{-3}$ in Fig.~\ref{fig:3}~(c).
Finally, just below a critical density $\rho_{\rm c}=\red{7.41}\times 10^{-4} \, \mathrm{fm}^{-3}$ \red{(which we shall derive below)}, as shown in Fig.~\ref{fig:3}~(d) where we set 
$\rho_\alpha=\red{7}\times 10^{-4}$ fm$^{-3}$,
the spectral weight is flattened in the low-momentum limit, which corresponds to the almost divergent effective mass.

To examine low-energy quasiparticle properties of a neutron immersed in the alpha condensate,
we perform the low-momentum expansion of $\Sigma_{s_z}(\red{\bm{p}},\omega)$.
First, one can find
\begin{equation}
\label{eq:17}
    \Sigma_{s_z}(\red{\bm{p}}=\bm{0},\omega)=0,
\end{equation}
because $\Sigma_{s_z}(\red{\bm{p}},\omega)$ is always proportional to $k^2$ due to the $p$-wave properties.
Accordingly, the $p$-wave polaron energy $E_{\rm P}$ is zero (i.e., $E_{\rm P}=\Sigma_{s_z}(\red{\bm{p}}=\bm{0},\omega)=0$).  
We expand $\Sigma_{s_z}(\red{\bm{p}},\omega)$ around $\red{\bm{p}}=\bm{0}$ and $\omega=\xi_{\red{\bm{p}}=\bm{0},\nu}=0$ as
\begin{align}
    \Sigma_{s_z}(\red{\bm{p}},\omega)&\simeq
    \Sigma_{s_z}(\bm{0},\xi_{\bm{0},\nu})\cr
    &
    \quad
    +\left.\frac{\partial \Sigma_{s_z}(\bm{0},\omega)}{\partial \omega}\right|_{\omega=\xi_{\bm{0},\nu}}
    (\omega-\xi_{\bm{0},\nu})\cr
    &
    \quad
    +\frac{1}{2}\left.\frac{\partial^2 \Sigma_{s_z}(\red{\bm{p}},\xi_{\bm{0},\nu})}{\partial \red{p}^2}\right|_{\red{p}=0}\red{p}^2.
\end{align}
In this way, we obtain the approximate form of polaronic neutron Green's function
\begin{align}
    G_{s_z}(\red{\bm{p}},\omega)
    &\simeq\frac{Z}{\omega_+-\frac{\red{p}^2}{2M_{{\rm eff}}}+i\Gamma_{\rm P}/2},
\end{align}
where $\Sigma_{s_z}(\bm{0},\xi_{\bm{0},\nu})=0$.

Here, we define the quasiparticle residue
\begin{equation}
    Z=\left[1-{\rm Re}\left.\frac{\partial \Sigma_{s_z}(\bm{0},\omega)}{\partial \omega}\right|_{\omega=0}
    \right]^{-1},
\end{equation}
and the inverse effective mass
\begin{align}
\label{eq:18}
    \frac{M_{\nu}}{M_{{\rm eff}}}=Z\left[1+M_\nu{\rm Re}\left.\frac{\partial^2 \Sigma_{s_z}(\red{\bm{p}},\xi_{\bm{0},\nu})}{\partial \red{p}^2}\right|_{\red{p}=0}\right].
\end{align}
The low-momentum expansion of $\Sigma_{s_z}(\red{\bm{p}},\omega)$ reads
\begin{equation}
\label{eq:sigma_lowk}
     \Sigma_{s_z}(\red{\bm{p}},\omega)
     =\frac{2}{3\red{\left(1+\frac{M_\nu}{M_\alpha}\right)^2}}
    \frac{g^2\rho_\alpha \red{p}^2}{\omega_+-E_{\Phi}-\Pi_{0}(\omega)} +O(\red{p}^4).
\end{equation}
We analytically obtain
    $\Pi_{0}(0)
    =-{M_{\rm r}g^2\Lambda^3}/{12\pi}$
and hence
\begin{equation}
    \left.\frac{\partial^2 \Sigma_{s_z}(\red{\bm{p}},0)}{\partial \red{p}^2}\right|_{\red{\bm{p}}=\bm{0}}
    \red{=-\frac{8\pi a_p}{M_{\nu}\left(1+\frac{M_\nu}{M_\alpha}\right)}\rho_\alpha.}
\end{equation}
Moreover, one can find
\begin{align}
  \frac{\partial\Sigma_{s_z}(\bm{0},\omega)}{\partial \omega}
  & =
    0, \ \rightarrow Z=1, \\
  {\rm Im}\Sigma_{s_z}(\bm{0},\omega)
  & =
    0, \ \rightarrow \Gamma_{\rm P}=0.
\end{align}
\red{Taking} $Z\rightarrow1$ in Eq.~\eqref{eq:18},
we obtain
\begin{align}
    \frac{M_{\nu}}{M_{\rm eff}}
    \red{=1+\frac{8\pi M_\alpha a_p\rho_\alpha}{{M_\nu}+{M_\alpha}}<1} \quad (a_p<0),
\end{align}
indicating the heavier effective mass compared to the bare one.
\red{This enhancement can be understood as the level repulsion between quasiparticle neutron and $^5$He branches where the quasiparticle neutron branch is pushed to lower energies by the coupling with a broad $^5$He resonance as found in Fig.~\ref{fig:3}. }
Eventually, formation of a bound dineutron in the alpha condensate is possible with the help of  
the induced interaction~\cite{PhysRevLett.121.013401}. 

If the alpha-condensation density increases further, the effective mass diverges at
\begin{align}
\label{eq:rho_c}
    \rho_{\rm c}
    \red{=\frac{1}{8\pi|a_p|}\left(1+\frac{M_\nu}{M_\alpha}\right)\simeq 7.41\times 10^{-4} \ {\rm fm}^{-3}.}
\end{align}
Beyond $\rho_{\rm c}$, the effective mass of a polaronic neutron becomes negative, indicating a breakdown of the low-momentum polaron picture in the rest frame of the homogeneous alpha condensate
and possibly a signature of self-localization of a polaronic neutron as discussed in Ref.~\cite{PhysRevA.92.033612,PhysRevA.97.053617,PhysRevB.109.235135}. 
This is in sharp contrast with $p$-wave Fermi polarons~\cite{PhysRevLett.109.075302,PhysRevA.100.062712}, of which a transition to a Feshbach molecular state was predicted to occur.
We also note that the negative effective mass was discussed in repulsive Fermi polarons as a precursor of ferromagnetic instability~\cite{PhysRevA.96.053609}.
Although we do not consider the alpha-alpha interaction and resulting quantum depletion explicitly, 
such effects would not change the result for the effective mass drastically. 
In fact,
it is reported that the effective mass is 
not strongly affected by
the boson-boson interaction in the case of $s$-wave Bose polarons~\cite{PhysRevA.88.053632}.

Strictly speaking, however, the alpha-alpha interaction should affect various properties of the medium itself even at $\rho\approx\rho_{\rm c}$. 
For example,
$4\rho_{\rm c}$ are not necessarily equal to the critical value of the total nucleon density because the depletion of the condensate due to quantum fluctuations occurs even at $T=0$. Indeed, such a depletion is predicted to originate from the inter-alpha interactions and the in-medium breakup of alpha particles themselves~\cite{clark2023alpha}.
Therefore, $\rho_{\rm c}$ may well be regarded as the critical density for the condensate component of alpha matter (not for the whole system).
Then, the corresponding total nucleon density could be much higher than $4\rho_{\rm c}$, which suggests that
the system is more like a liquid state rather than a gas state as usually assumed in earlier investigations~\cite{PhysRevC.80.064326}.

\begin{figure}[t]
    \centering
    \includegraphics[width=7cm]{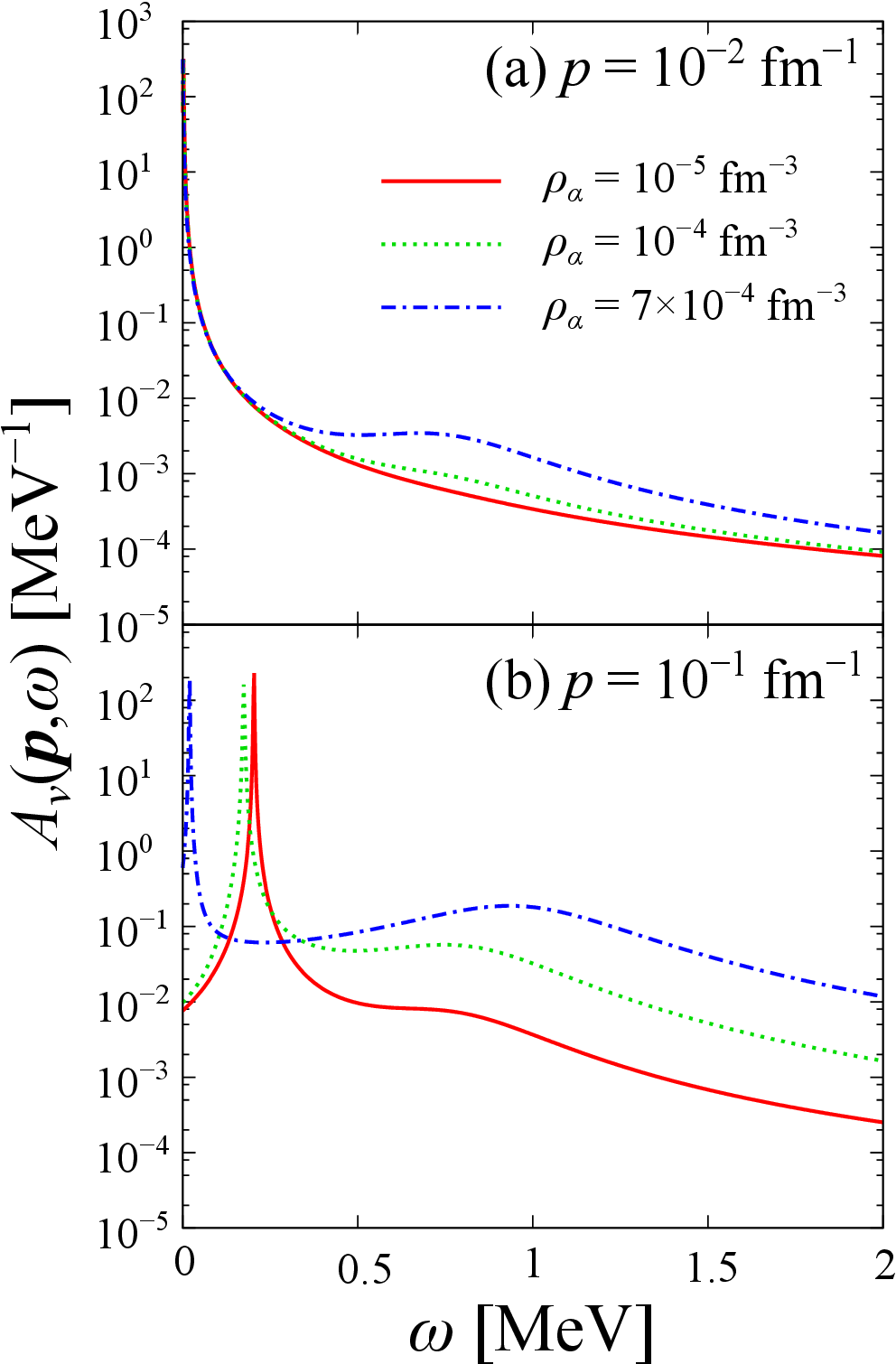}
    \caption{Polaronic neutron spectral weight $A_{\red{\nu}}(\red{\bm{p}},\omega)$ with fixed momenta [(a) $\red{p}=10^{-2} \ {\rm fm}^{-1}$ and (b) $\red{p}=10^{-1} \ {\rm fm}^{-1}$] in the alpha condensate. }
    \label{fig:4}
\end{figure}

To see more detailed structure of the neutron spectral weight,
we plot $A_{\red{\nu}}(\red{\bm{p}},\omega)$ at fixed momenta and different $\rho_\alpha$ in Fig.~\ref{fig:4}, where $\red{p}=10^{-2} \, \mathrm{fm}^{-1}$ and $\red{p}=10^{-1} \, \mathrm{fm}^{-1}$ are taken in the panels (a) and (b), respectively.
At a sufficiently small momentum as shown in the panel (a) of Fig.~\ref{fig:4},
a stable polaron peak can be found even near the critical density.
\red{On the other hand, at larger $\rho_\alpha$, one can find a small peak of $A_{\red{\nu}}(\bm{p},\omega)$ around the $^5$He resonance ($E_{\rm res}\simeq 0.93$~MeV), implying the excited branch associated with the level repulsion between quasiparticle neutron and $^5$He branches.
Because the width of the $^5$He resonance ($\simeq 0.6$ MeV) is substantially broad in the present energy scale~\cite{PhysRevC.106.045807}, however, this peak is also largely broadened.
At larger frequencies, the neutron spectral weight monotonically decreases.}
Building the high-frequency limit of Eq.~\eqref{eq:pi},
\begin{equation}
    \Pi_0(\omega\rightarrow\infty)\rightarrow -i\frac{M_{\rm r}g^2\Lambda^4}{6\pi\sqrt{2M_{\rm r}\omega}},
\end{equation}
into ${\rm Im}\Sigma_{s_z}(\red{\bm{p}},\omega\rightarrow\infty)$ as
\begin{align}
  &  {\rm Im}\Sigma_{s_z}(\red{\bm{p}},\omega\rightarrow\infty)
  \rightarrow \frac{2}{3}\frac{g^2\rho_\alpha {k}^2{\rm Im}\Pi_0(\omega\rightarrow\infty)}{\omega^2}+O(k^4)\cr
    &
    \quad
    \quad
    \quad
    \quad
    \equiv -\frac{M_{\rm r}g^4\Lambda^4\rho_\alpha \red{p}^2}{9\pi
    \sqrt{2M_{\rm r}}\red{\left(1+\frac{M_\nu}{M_\alpha}\right)^2}\omega^{5/2}}+O(\red{p}^4),
\end{align}
we obtain the high-frequency tail of the spectral weight at low momenta as
\begin{equation}
\label{eq:hftail}
    A_{\red{\nu}}(\red{\bm{p}},\omega\rightarrow\infty)\rightarrow
    \frac{M_{\rm r}g^4\Lambda^4\rho_\alpha \red{p}^2}{9\pi^2\sqrt{2M_{\rm r}}
    \red{\left(1+\frac{M_\nu}{M_\alpha}\right)^2}
    \omega^{9/2}}+O(\red{p}^4),
\end{equation}
indicating that the high-frequency tail is proportional to $\rho_\alpha$ as well as $\red{p}^2$.
Indeed, Fig.~\ref{fig:4}(a) shows the enhancement of the high-frequency spectra with increasing $\rho_\alpha$.
As shown in Fig.~\ref{fig:4}(b), at a larger momentum,
the polaron peak is shifted towards high frequencies 
in such a way as to follow the dispersion relation $\red{p}^2/2M_{\rm eff}$ at relatively small $\rho_\alpha$.
We remark in passing that in a manner that is consistent with the $p$ dependence of Eq.~\eqref{eq:hftail},
the high-frequency contribution in the panel (b) is significantly larger than
that in the panel (a).

\subsection{Dineutrons in the alpha condensate}
Here, we show that a dineutron can be a bound state due to the enhanced $M_{\rm eff}$ together with the $^1S_0$ neutron-neutron interaction $U_{2\nu}(k,k')$.
A similar binding mechanism of two alpha particles in dilute neutron matter has been examined theoretically~\cite{PhysRevC.104.065801}. 
On the basis of the results of polaronic neutron spectra shown in Fig.~\ref{fig:3},
we employ the approximated form of the polaronic neutron  propagator in dilute alpha matter as
\begin{align}
    \red{
    \tilde{G}_{s_z}(\bm{p},\omega)\simeq \frac{\theta(k_{\rm c}-p)}{\omega_+-{p^2}/{2M_{\rm eff}}}+
    \frac{\theta(p-k_{\rm c})}{\omega_+-{p^2}/{2M_{\nu}},}
    }
    \label{eq:39v2}
\end{align}
where we have assumed that the low-energy polaron state is valid up to some momentum cutoff $k_{\rm c}$.
\red{Eq.~\eqref{eq:39v2} imitates two branches induced by the avoided crossing of neutron quasiparticle and $^5$He resonance branches except for the region near $\rho_{\alpha}=\rho_{\rm c}$.}
\red{In this work, we specifically focus on the role of the enhanced neutron effective mass at low momenta below $k_{\rm c}$ found in Fig.~\ref{fig:3}.}
The precise value of $k_{\rm c}$, which is controlled by the properties of the alpha condensate and the neutron-alpha interaction, remains to be known and hence will be taken arbitrarily in the range of $0.1 \ {\rm fm}^{-1}\lesssim k_{\rm c}\lesssim 10 \ {\rm fm}^{-1}$.
\red{
While the spectral broadening due to the coupling with the $^5$He resonance should also be important above $k=k_{\rm c}$,
we consider two bare neutrons in a virtual state there for simplicity because we are interested in the ground state of the two-neutron system in the alpha condensate, where the low-momentum properties, namely, enhanced effective masses play a pivotal role.
We note that the higher-momentum properties of neutrons would be important when neutrons undergo the Fermi degeneracy at finite excess neutron densities.
}

The two-neutron $T$-matrix $T_{2\nu}(k,k',\red{\Omega})$ 
in the $^1S_0$ channel is given by
\begin{widetext}
  \begin{equation}
    T_{2\nu}(k,k',\red{\Omega})
      =U_{2\nu}(k,k')
      \red{+i
      \sum_{\bm{q}}\int_{-\infty}^{\infty}\frac{d\omega}{2\pi}
      U_{2\nu}(k,q)
    \tilde{G}_{+1/2}(\bm{q},\omega+\Omega)
    \tilde{G}_{-1/2}(-\bm{q},-\omega)
    T_{2\nu}(q,k',\Omega)}
\end{equation}
\end{widetext}
Using the separable $^1S_0$ neutron-neutron interaction $U_{2\nu}(k,k')=U_0\chi_k\chi_{k'}$ with the form factor $\chi_k$ 
to be specified below, we obtain 
$T_{2\nu}(k,k';\red{\Omega})=t_{2\nu}(\red{\Omega})\chi_{k}\chi_{k'}$
with
\begin{equation}
    t_{2\nu}(\red{\Omega})
    =\frac{U_0}{1-U_0\Xi(\red{\Omega})},
\end{equation}
where the polarization function $\Xi$ reads
\red{
\begin{align}
\label{eq:35}
    \Xi(\Omega)
    &=\sum_{\bm{q}}\frac{\chi_q^2\theta(k_{\rm c}-q)}{\Omega_+-q^2/M_{\rm eff}}
    +\sum_{\bm{q}}\frac{\chi_q^2\theta(q-k_{\rm c})}{\Omega_+-q^2/M_\nu},
\end{align}}
\red{with $\Omega_+=\Omega+i\delta$.}
The dineutron binding energy $E_{\red{2\nu}}$ can be obtained from
\begin{equation}
    1-U_0\Xi(\Omega=-E_{\red{2\nu}})=0.
\end{equation}

\begin{figure}[t]
    \centering
    \includegraphics[width=8cm]{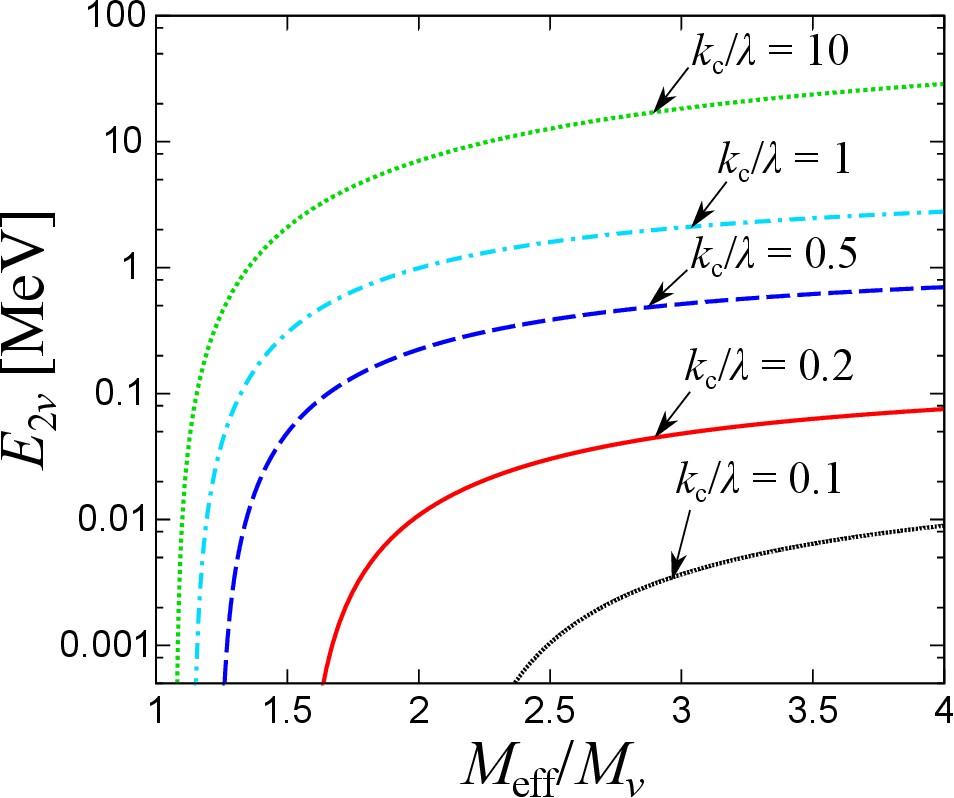}
    \caption{Dineutron binding energy $E_{\red{2\nu}}$ in dilute alpha matter as a function of the effective mass $M_{\rm eff}$ of a polaronic neutron with various values of the cutoff parameter $k_{\rm c}$ divided by $\lambda=0.765 \, \mathrm{fm}^{-1}$.}
    \label{fig:5}
\end{figure}

Practically, we use
    $\chi_q=1/{\sqrt{1+(q/\lambda)^2}}$.
The parameters $U_0$ and $\lambda$ are determined by the $^1S_0$ scattering length $a_{2\nu }=-18.5 \, \mathrm{fm} $ and $r_{2\nu}=2.8 \, \mathrm{fm} $ as~\cite{PhysRevA.97.013601}
\begin{align}
  U_0
  & =
    \frac{4\pi a_{2\nu}}{M_\nu}\frac{1}{1-a_{2\nu}\lambda}, \quad
  \lambda
  =
    \frac{1}{r_{2\nu}}\left(1+\sqrt{1-\frac{2r_{2\nu}}{a_{2\nu}}}\right).
    \label{eq:39}
\end{align}
Moreover, one can perform the momentum summation in 
Eq.~\eqref{eq:35}
as
\begin{widetext}    
\begin{align}
    \Xi(-E_{\red{2\nu}})
        &=-\frac{M_\nu\lambda^2}{2\pi^2}
    \left[\frac{M_{\rm eff}}{M_\nu}\frac{\lambda\tan^{-1}\left(\frac{k_{\rm c}}{\lambda}\right)-\sqrt{M_{\rm eff}E_{\red{2\nu}}}\tan^{-1}\left(\frac{k_{\rm c}}{\sqrt{M_{\rm eff}E_{\red{2\nu}}}}\right)}{\lambda^2-M_{\rm eff}E_{\red{2\nu}}}\right.\cr
    &\left.
    \quad+\frac{\pi}{2}\frac{1}{\lambda+\sqrt{M_\nu E_{\red{2\nu}}}}
    -\frac{\lambda\tan^{-1}\left(\frac{k_{\rm c}}{\lambda}\right)-\sqrt{M_{\nu}E_{\red{2\nu}}}\tan^{-1}\left(\frac{k_{\rm c}}{\sqrt{M_{\nu}E_{\red{2\nu}}}}\right)}{\lambda^2-M_{\nu}E_{\red{2\nu}}}\right].
\end{align}
In this way, the bound-state equation reads
\begin{align}
\label{eq:dineutron}
    1-\frac{1}{a_{2\nu}\lambda}
    &=\frac{2}{\pi}
    \left[
    \frac{M_{\rm eff}}{M_\nu}
    \frac{\tan^{-1}\left(\frac{k_{\rm c}}{\lambda}\right)-\sqrt{\frac{M_{\rm eff}}{M_\nu}\frac{M_{\nu}E_{\red{2\nu}}}{\lambda^2}}\tan^{-1}\left(\frac{k_{\rm c}}{\lambda}\sqrt{\frac{M_\nu}{M_{\rm eff}}\frac{\lambda^2}{M_{\nu}E_{\red{2\nu}}}}\right)}{1-\frac{M_{\rm eff}}{M_\nu}\frac{M_\nu E_{\red{2\nu}}}{\lambda^2}}\right.\cr
    &\left.
    \quad+\frac{\pi}{2}\frac{1}{1+\sqrt{\frac{M_\nu E_{\red{2\nu}}}{\lambda^2}}}
    -\frac{\tan^{-1}\left(\frac{k_{\rm c}}{\lambda}\right)-\sqrt{\frac{M_{\nu}E_{\red{2\nu}}}{\lambda^2}}\tan^{-1}\left(\frac{k_{\rm c}}{\lambda}\sqrt{\frac{\lambda^2}{M_{\nu}E_{\red{2\nu}}}}\right)}{1-\frac{M_{\nu}E_{\red{2\nu}}}{\lambda^2}}\right].
\end{align}
\end{widetext}
The numerical results for $E_{\red{2\nu}}$ that can be obtained from Eq.~\eqref{eq:dineutron} are shown in Fig.~\ref{fig:5}.
These results indicate that dineutrons can be bound due only to the effective-mass acquisition of polaronic neutrons and the direct $^1S_0$ neutron-neutron interaction, even without the help of the medium-induced interaction. 
In this case, there is a threshold effective mass $M_{\rm eff,c}$ for the presence of a bound dineutron, which can be obtained by taking $E_{\red{2\nu}}=0$ in Eq.~\eqref{eq:dineutron} as
\begin{equation}
\label{eq:meffc}
     \frac{M_{\rm eff, c}}{M_\nu}=
     1-\frac{\pi}{2a_{2\nu}\lambda\tan^{-1}\left(\frac{k_{\rm c}}{\lambda}\right)}.
\end{equation}

Here, one needs to examine the cutoff-parameter dependence of $E_{\red{2\nu}}$ and $M_{\rm eff, c}$.
As shown in Fig.~\ref{fig:5} where several different values of $k_{\rm c}/\lambda$ are taken with $\lambda$ fixed at $0.765$ fm$^{-1}$ [see Eq.~\eqref{eq:39}], $E_{\red{2\nu}}$ increases with $k_{\rm c}$.  This is because for larger $k_{\rm c}$, each neutron more often behaves like a polaronic massive particle; according to Eq.~\eqref{eq:dineutron}, for $k_{\rm c}\gg\lambda$ where $\tan^{-1}(k_{\rm c}/\lambda)\simeq\pi/2$, $E_{\red{2\nu}}$ is close to its upper limit $\lambda^2/M_{\rm eff}[(M_{\rm eff}/M_\nu)/(1-1/a_{2\nu}\lambda)-1]^2$,
while for $k_{\rm c}\ll\lambda$ where $\tan^{-1}(k_{\rm c}/\lambda)\simeq k_{\rm c}/\lambda$, the condition for $E_{\red{2\nu}}>0$ is sensitive to the value of $k_{\rm c}$.
Also, $M_{\rm eff, c}$ tends to decrease with increasing $k_{\rm c}$, as is clear from Eq.~\eqref{eq:meffc}.
While the notable $k_{\rm c}$ dependence of $E_{\red{2\nu}}$ can be found in Fig.~\ref{fig:5},
it would be fair to employ $k_{\rm c}/\lambda=0.2$ (i.e., $k_{\rm c}\simeq 0.15 \, \mathrm{fm}^{-1}$), below which
the polaronic spectral weight exhibits a sharp quasiparticle peak as
shown in Figs.~\ref{fig:3}(a) and (b).

\begin{figure}[t]
    \centering
    \includegraphics[width=8cm]{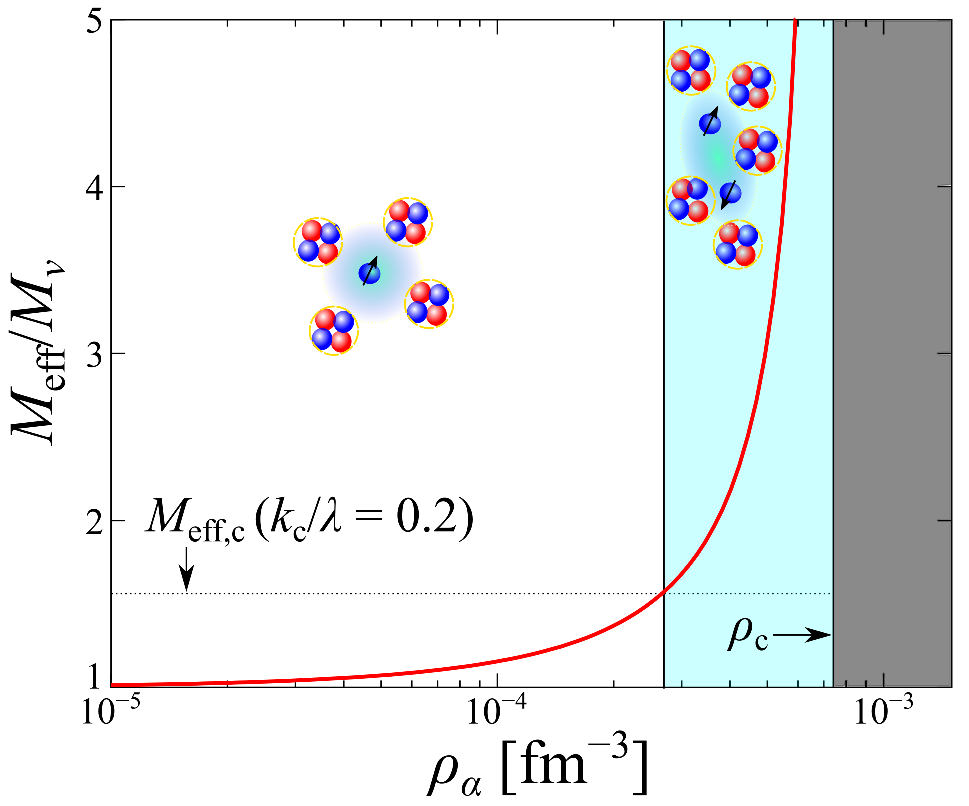}
    \caption{Effective mass $M_{\rm eff}$ of a polaronic neutron plotted as a function of the alpha condensation density $\rho_\alpha$.  The horizontal dotted line shows the critical effective mass $M_{\rm eff,c}$ for the presence of a bound dineutron in the case of $k_{\rm c}/\lambda=0.2$; for momenta below this $k_{\rm c}$, a sharp peak can be seen in the calculated neutron spectral weight.
    $\rho_{\rm c}$ is the critical alpha condensation density given by Eq.~\eqref{eq:rho_c}. In the shaded region beyond $\rho=\rho_{\rm c}$, the polaronic neutron experiences divergence of the effective mass, indicating the breakdown of the low-momentum polaron picture in the rest frame of the alpha condensate and possibly the resultant self-localization~\cite{PhysRevA.92.033612,PhysRevA.97.053617,PhysRevB.109.235135}.}
    \label{fig:6}
\end{figure}

In Fig.~\ref{fig:6}, we show the effective mass $M_{\rm eff}$ of a polaronic neutron as a function of $\rho_\alpha$.
For comparison, we also show the critical effective mass $M_{\rm eff, c}$ for the formation of a bound dineutron in the alpha condensate in the case of $k_{\rm c}/\lambda=0.2$.
In this case, a bound dineutron starts to appear around $\rho_\alpha\simeq \red{3}\times 10^{-4} \, \mathrm{fm}^{-3}$,
where polaronic neutrons remain stable and become sufficiently massive.  
Such kind of formation of bound dineutrons 
might be relevant to the alpha and dineutron clustering structure of neutron-rich light nuclei such as $^8$He~\cite{PhysRevLett.131.242501}
and $^{10}$Be~\cite{PhysRevC.61.044306,dan2021description}.
While the condensation density $\rho_\alpha$ is substantially lower than the normal nuclear density, as already discussed in the previous subsection, 
$\rho_\alpha$ could be substantially lower than the total alpha density including the effects of quantum depletion.  Also,
the dineutron formation in the alpha condensate via the neutron-alpha $p$-wave coupling is similar to the structure of two-neutron halo nuclei such as
$^{6}$He and $^{11}$Li.
To take a closer look at this similarity, the $J^\pi=1/2^{-}$ channel as neglected in the present study would have to be taken into account.

\section{Summary}
\label{sec:4}
In this paper, we have investigated the properties of a polaronic neutron in dilute alpha matter; the Bose polaron picture adopted here has been extensively examined in cold-atomic physics.
We have shown that each polaronic neutron undergoes a large enhancement of the effective mass, a vanishing decay, and a constant quasiparticle residue, because of the resonant $p$-wave coupling with the alpha condensate.
Such effective mass enhancement helps to form bound dineutrons, together with the $^1S_0$ neutron-neutron attraction, which is not sufficiently strong to induce the dineutron binding in a vacuum.

To examine more quantitative properties of bipolaronic dineutrons, effects of spectral broadening, finite neutron density and temperature, the alpha-mediated interaction between polarons, and larger multi-nucleon clusters will have to be considered. 
In particular, while we have addressed bound dineutrons only via the increased effective mass and the residual $^1S_0$ neutron-neutron interaction, the alpha-mediated interaction would further induce a larger dineutron binding energy 
as in the case of bipolarons
~\cite{naidon2018two,PhysRevLett.121.013401,Panochko_2021,PhysRevLett.129.233401,paredes2024perspective,nakano2024two}.

The extension to the medium system with quartet condensation or quartet BCS models~\cite{PhysRevLett.80.3177,PhysRevC.85.061303,sen2011unified,Sandulescu2015Phys.Lett.B751_348,BARAN2020135462,PhysRevC.105.024317} would also be a fascinating direction.
Moreover, the enhanced effective mass
may lead to multi-nucleon clusters~\cite{PhysRevLett.116.052501,FAESTERMANN2022136799,duer2022observation}.

\begin{acknowledgments}
  H.~T. is grateful to S. Furusawa for useful discussion on supernova matter.
  This work is supported by Grants-in-Aid for Scientific
  Research provided by JSPS through
  Nos.~JP18H05406,
  JP22H01158,
  JP22H01214,
  JP22K13981,
  JP22K20372, 
  JP22K25864,
  JP23H01845,
  JP23H04526,
  JP23H01845,
  JP23K01845,
  JP23K03426,
  JP23K22485,
  JP23K25864,
  JP24K17057,
  and 
  JP24K06925.
  T.~N.~acknowledges 
  the RIKEN Special Postdoctoral Researcher Program.
\end{acknowledgments}

\bibliography{reference.bib}
\end{document}